\documentclass[]{interact}

\usepackage[caption=false]{subfig}

\usepackage[natbibapa,nodoi]{apacite}
\usepackage{amsmath}
\usepackage{hyperref}
\usepackage{url}
\usepackage{color}
\usepackage{graphicx}
\usepackage{booktabs}
\usepackage{lipsum}
\usepackage[super]{nth}
\usepackage{amssymb}
\usepackage{multirow}
\usepackage{cite}

\DeclareMathOperator*{\argmax}{arg\,max}

\begin{document}


\title{Automatic Melody Harmonization with Triad Chords: \\
A Comparative Study}
%
%

\author{Yin-Cheng Yeh, Wen-Yi Hsiao, Satoru Fukayama, Tetsuro Kitahara, \\ Benjamin Genchel, Hao-Min Liu, Hao-Wen Dong, Yian Chen, Terence Leong, and Yi-Hsuan Yang}
\thanks{Yeh, Hsiao, Liu, Dong and Yang are with Academia Sinica, Taiwan (\{ycyeh, wayne391, paul115236, salu133445, yang\}@citi.sinica.edu.tw); Fukayama is with National Institute of Advanced Industrial Science and Technology, Japan (satoru s.fukayama@aist.go.jp); Kitahara is with Nihon University, Japan (kitahara@chs.nihon-u.ac.jp); Genchel is with Georgia Institute of Technology, USA (benjiegenchel@gmail.com); Chen and Leong are with KKBOX Inc., Taiwan (annchen@kkbox.com, terenceleong@kkboxgroup.com)}

\maketitle

\begin{abstract}
Several prior works have proposed various methods for the task of automatic melody harmonization, in which a model aims to generate a sequence of chords to serve as the harmonic accompaniment of a given multiple-bar melody sequence.
In this paper, we present a comparative study evaluating and comparing the performance of a set of canonical approaches to this task, including a template matching based model, a 
hidden Markov based model, a genetic algorithm based model, and two deep learning based models. 
The evaluation is conducted on a dataset of 9,226 melody/chord pairs we newly collect for this study, considering up to 48 triad chords, using a standardized training/test split.
We report the result of an objective evaluation using six different metrics and a subjective study with 202 participants.

\end{abstract}


\section{Introduction}\label{sec:introduction}

\begin{figure}[]
    \centering
    \includegraphics[width=.6\columnwidth]{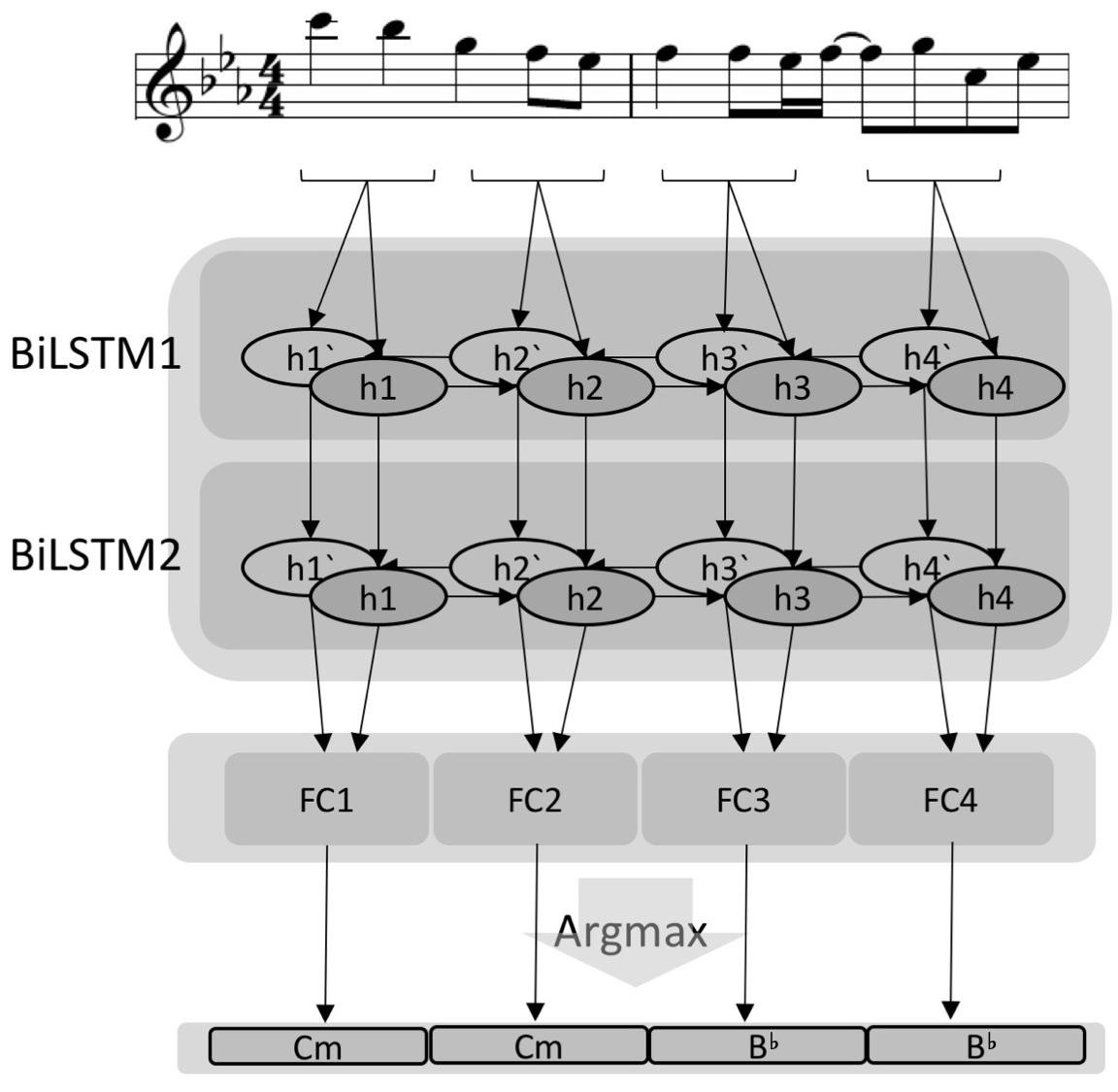}
    \caption{Diagram of the slightly modified version of the bidirectional long short-term memory network (BiLSTM) based model \citep{lim17} for melody harmonization. The input to the model is a melody sequence. With two layers of BiLSTM and one fully-connected (FC) layer, the model generates as output a sequence of chord labels (e.g., Cm or B chords), one for each half bar. See Section \ref{sec:bilstm} for details.}
    \label{fig:BiLSTM_Model}
\end{figure}

Automatic melody harmonization, a sub-task of automatic music generation \citep{fernandez13jair}, refers to the task of creating computational models that can generate a harmonic accompaniment for a given melody \citep{chuan07,simon08}. 
Here, the term harmony, or harmonization, is used to refer to chordal accompaniment, where an accompaniment is defined relative to the melody as the supporting section of the music. 
Figure \ref{fig:BiLSTM_Model} illustrates the inputs and outputs for a melody harmonization model.

Melody harmonization is a challenging task as there are multiple ways to harmonize the same melody; 
what makes a particular harmonization pleasant is subjective, and often dependent on musical genre and other contextual factors. Tonal music, which encompasses most of Western music, defines specific motivic relations between chords based on scales such as those defined in functional harmony \citep{riemann1893}.
While these relations still stand and are taught today, their application towards creating pleasant music often depends on subtleties, long term dependencies and cultural contexts which may be readily accessible to a human composer, but very difficult to learn and detect for a machine. While a particular harmonization may be deemed technically correct in some cases, it can also be seen as
uninteresting in a modern context.

There have been several efforts made towards this task in the past \citep{makris2016automatic}.
Before the rise of deep learning, the most actively employed approach is based on hidden Markov models (HMMs). 
For example, \citet{paiement2006pmh} proposed a tree-structured HMM that allows for learning the non-local dependencies of chords, and encoded probabilities for chord substitution taken from psycho-acoustics. They additionally presented a novel representation for chords that encodes relative scale degrees rather than absolute note values, and included a sub-graph in their model specifically for processing it. \citet{tsushima17ismir} similarly presented a hierarchical tree-structured model combining probabilistic context-free grammars (PCFG) for chord symbols and HMMs for chord rhythms. \citet{temperley2009unified} 
presented a statistical model that would generate and analyze music along three sub-structures: metrical structure, harmonic structure, and stream structure. In the generative portion of this model, a metrical structure defining the emphasis of beats and sub-beats is first generated, and then harmonic structure and progression are generated conditioned on that metrical structure. 

There are several previous works which attempt to formally and probabilistically analyze tonal harmony and harmonic structure. 
For example, \citet{rohrmeier2008statistical} applied a number of statistical techniques to harmony in Bach chorales in order to uncover a proposed underlying harmonic syntax that naturally produces common perceptual and music theoretic patterns including functional harmony. \citet{jacoby2015information} attempted to categorize common harmonic symbols (scale degrees, roman numerals, or sets of simultaneous notes) into higher level functional groups, seeking underlying patterns that produce and generalize functional harmony. \citet{tsushima2018generative} used unsupervised learning in training generative HMM and PCFG models for harmonization, showing that the patterns learned by these models match the categorizations presented by functional harmony.

More lately, people have begun to explore the use of deep learning for a variety of music generation tasks \citep{briot17survey}. For melody harmonication, \citet{lim17} proposed a model that employed two bidirectional long short-term memory (BiLSTM) recurrent layers \citep{hochreiter97LSM} and one fully-connected layer to learn the correspondence between pairs of melody and chord sequences. The model architecture is depicted in Figure \ref{fig:BiLSTM_Model}. According to the experiments reported in \citep{lim17}, this model outperforms a simple HMM model and a more complicated DNN-HMM model \citep{hinton12spm} for melody harmonization with major and minor triad chords.

Moreover, melody harmonization is also relevant to four-part chorale harmonization~\citep{ebcioglu1998chorale,allan2005chorale,huang17ismir,hadjeresP17icmlr} and accompaniment generation~\citep{chuan07,simon08,musegan}. Although these three tasks share the same input, monophonic melodies, they have distinct outputs---lead sheets, four-part chorales and full arrangements, respectively. Another relevant topic is harmonic analysis~\citep{raphael2004harmonic,tonalDist,temperley2009unified,dehaas2014functional,chen18ismir}. Despite having similar input and output spaces, melody harmonization is fundamentally different from harmonic analysis. On one hand, harmonic analysis takes polyphonic music as inputs, while melody harmonization takes monophonic melodies as inputs and is considered more difficult as less harmonic information is given. On the other hand, there is in general a correct answer for harmonic analysis, while there are no strict answers for melody harmonization.

We note that, while many new models are being proposed for melody harmonization, at present there is no comparative study evaluating a wide array of different approaches for this task, using the same training set and test set. 
Comparing models trained on different training sets is problematic as it is hard to have a standardized definition of improvement and quality.
Moreover, as there is to date no standardized test set for this task, it is hard to make consistent comparison between different models.

In this paper, we aim to bridge this gap with the following three contributions:
\begin{enumerate}    
    \item We implement in total five melody harmonization models that span a number of canonical approaches to the task, including template matching, hidden Markov model (HMM) \citep{simon08}, genetic algorithm (GA) \citep{kitahara2018}, and two variants of deep recurrent neural network (RNN) models \citep{lim17}. 
    We then present a comparative study comparing the performance of these models. 
    To our best knowledge, a comparative study that considers such a diverse set of approaches for melody harmonization using a standardized dataset has not been attempted before.
    
    As we follow fairly faithfully the implementation proposed in the original publications, these models differ in terms of not only the model architectures but also the employed features. Therefore, we have to admit that our study cannot decouple the effects of the model architectures and the features. Yet, we note that the comparison of the first four models is an architecture-vs-architecture comparison, while the comparison of the two RNN models is a feature-vs-feature comparison.
    \item We compile a new dataset, called the Hooktheory Pianoroll Triad Dataset (HTPD3), to evaluate the implemented models over well-annotated
    lead sheet samples of music. 
    A lead sheet is a form of musical notation that specifies the essential elements of a song---the melody, harmony, and where present, lyrics \citep{liu18icmla}. HTPD3 provides melody lines and accompanying chords 
    specifying both chord symbol and harmonic function useful for our study. 
    We consider 48 triad chords in this study, including major, minor, diminished, and augmented triad chords.

    We use the same training split of HTPD3 to train the implemented models and evaluate them on the same test split.
    \item We employ six objective metrics for evaluating the performance of melody harmonization models. 
    These metrics consider either the distribution of chord labels in a chord sequence, or how the generated chord sequence fits with the given melody. In addition, we conduct an online user study and collect the feedback from 202 participants around the world to assess the quality of the generated chordal accompaniment. 
\end{enumerate}

We discuss the findings of comparative study, hoping to gain insights into the strength and weakness of the evaluated methods.
Moreover, we show that incorporating the idea of functional harmony \citep{chen18ismir} 
while harmonizing melodies greatly improves the result of the model presented by \citep{lim17}.




In what follows, we present in Section \ref{sec:models} the models we consider and evaluate in this comparative study. Section \ref{sec:dataset} provides the details of the HTPD3 dataset we build for this study, and Section \ref{sec:metrics} the objective metrics we consider. Section \ref{sec:eval} presents the setup and result of the study. We discuss the findings and limtiations of this study in Section \ref{sec:discussion}, and then conclude the paper in Section \ref{sec:conclusion}.

\section{Automatic Melody Harmonization Models}
\label{sec:models}

A melody harmonization model takes a melody sequence of $T$ bars as input and generates a corresponding 
chord sequence as output. \emph{Chord Sequence} is defined here as a series of chord labels $Y=y_1,y_2,\dots,y_M$, where $M$ denotes the length of the sequence.
In this work, each model predicts a chord label for every half bar, i.e. $M=2T$.
Each label $y_j$ is chosen
from a finite chord vocabulary $\mathcal{C}$.
To reduce the complexity of this task, we consider here only the triad chords, i.e., chords composed of three notes. 
Specifically, we consider major, minor, diminished, and augmented triad chords, all in root position.
We also consider \emph{No Chord} (N.C.), or rest, so the size of the chord vocabulary is $|\mathcal{C}|=49$.
\emph{Melody Sequence} is a time-series of monophonic musical notes in MIDI 
format. We compute a sequence of features as $X=\mathbf{x}_1,\mathbf{x}_2,\dots,\mathbf{x}_{N}$ to represent
the melody and use them as the inputs to our models. Unless otherwise specified, we set $N=M$, computing a feature vector for each half bar. 

Given a set of melody and corresponding chord sequences, a melody harmonization model $f(\cdot)$ can be trained by minimizing
the loss computed between the ground truth $Y_*$ and the model output $\hat{Y}_*=f(X_*)$, where $X_*$ is the input melody.


We consider three non-deep learning based and two deep learning based models in this study.
While the majority are adaptation of existing methods, one (deep learning based) is a novel method which we introduce in this paper (see Section \ref{sec:models:ours}). 
All models are carefully implemented and trained 
using the training split of HTPD3.
We present the technical details of these models below.


\subsection{Template Matching-based Model}
\label{sec:models_rule}

This model is based on an early work on audio-based chord recognition \citep{fujishima99}. 
The model segments training melodies into half-bars, and constructs a \emph{pitch profile} for each segment. The chord label for a new segment is then selected based on the label for the training segment whose pitch profile it most closely matches. 
\textcolor{black}{When there is more than one possible chord template that has the highest matching score, we choose a chord randomly based on uniform distribution among the possibilities.}
We refer to this model as \textit{template matching-based} as the underlying method \textcolor{black}{compares the profile of a given melody segment with those of the \emph{template} chords.}



We use Fujishima's \emph{pitch class profile} (PCP) \citep{fujishima99}
as the pitch profile representing respectively the melody and chord for each half-bar. A PCP is a 12-dimensional feature vector $\mathbf{x} \in [0,1]^{12}$ where each element corresponds to the activity of a pitch class. The PCP for each of the $|\mathcal{C}|$ chord labels is constructed by setting the elements corresponding to the pitch classes that are part of the chord to one, and all the others to zero. Because we consider only triad chords in this work, there will be exactly three one's in the PCP of a chord label for each half bar.
The PCP for melody is constructed similarly, but additionally considering the duration of notes. Specifically, the activity of the $k$-th pitch class, i.e., $x_k \in [0,1]$,  is set by the ratio of time the pitch class is active during the corresponding half bar. 



The result of this model are more conservative by design, featuring intensive use of chord tones. And, this model sets the chord label independently for each half bar, without considering the neighboring chord labels, or the chord progression over time.

We note that, to remove the effect of the input representations on the harmonization result, we use PCP as the model input representation for all the other models we implement for melody harmonizationm.

\subsection{HMM-based Model}
\label{sec:hmmmodel}

HMM is a probabilistic framework for modeling sequences with latent or hidden variables.
Our HMM-based harmonization model regards chord labels as latent variables and estimates the most likely chord sequence for a given set of melody notes.
Unlike the template matching-based model, this model considers the relationship between neighboring chord labels. HMM-based models similar to this one were widely used in chord generation and melody harmonization research before the current era of deep learning   \citep{simon08,raczynski13}.


We adopt a simple HMM architecture employed in \citep{lim17}. This model makes the following assumptions:
\begin{itemize}
    \item[1.] The observed melody sequence $X=\mathbf{x}_1,\dots,\mathbf{x}_M$ is statistically biased due to the hidden chord sequence $Y=y_1,\dots,y_M$, which is to be estimated.
    \item[2.] $\mathbf{x}_m$ depends on only $y_m$, $\forall m \in [1,M]$.
    \item[3.] $y_m$ depends on only $y_{m-1}$, $\forall m \in [2,M]$.
\end{itemize} 
The task is to estimate the  most likely hidden sequence $\hat{Y}=\hat{y}_1,\dots,\hat{y}_M$ given $X$. This amounts to maximizing the posterior probability:
\begin{eqnarray}
   \hat{Y} &=& \argmax_{Y} P(Y|X) ~=~ \argmax_{Y} P(X|Y)P(Y) \nonumber \\
    &=& \argmax_{Y} \prod_{m=1}^{M} P(\mathbf{x}_m | y_m ) P(y_m | y_{m-1}) \label{eq:prod}, 
\end{eqnarray}
where $P( y_1 | y_0)$ is equal to $P(y_1)$.
The term 
$P(\mathbf{x}_m | y_m )$ is also called the emission probability, and the term $P(y_m | y_{m-1})$ is called the transition probability. This optimization problem can be solved by the Viterbi algorithm \citep{viterbi}. 

Departing from the HMM in \citep{lim17}, our implementation uses the PCPs described in Section \ref{sec:models_rule} to represent melody notes, i.e., to compute $\mathbf{x}_m$. 
Accordingly, we use multivariate Gaussian distributions to model the emission probabilities, as demonstrated by Fujishima \citep{sheh03}.
For each chord label, we set the covariance matrix of the corresponding Gaussian distribution to be a diagonal matrix, 
and calculate the mean and variance for each dimension from the PCP features of melody segments that are associated with that chord label in the training set.

To calculate the transition probabilities, we count the number of transitions between successive chord labels (i.e., bi-grams), then normalize those counts to sum to one for each preceding chord label. A uniform distribution is used when there is no bi-gram count for the preceding chord label. To avoid zero probabilities, we smooth the distribution by interpolating 
$P(y_m | y_{m-1})$ with the prior probability $P(y_m)$ as follows,
\begin{equation}
    P'(y_m | y_{m-1}) = (1-\beta)  P(y_m) + \beta  P(y_m | y_{m-1}) \,,
\end{equation}
yielding the revised transition probability $P'(y_m | y_{m-1})$.
The hyperparameter $\beta$  
is empirically set to 0.08 via experiments on a random 10\% subset of the training set.


\subsection{Genetic Algorithm (GA)-based Model}
\label{sec:gamodel}

A GA is a flexible algorithm that generally maximizes an objective function or {\em fitness function}. GAs have been used for melody generation and harmonization in the past \citep{phon99,leo2016}, justifying their inclusion in this study. A GA can be used in both rule-based and probabilistic approaches. In the former case, we need to design a rule set of what conditions must be satisfied for musically acceptable melodies or harmonies---the fitness function is formulated based on this rule set. In the latter, the fitness function is formulated based on statistics of a data set. 

Here, we design a GA-based melody harmonization model by adapting the GA-based melody generation model proposed by \citep{kitahara2018}.
Unlike the other implemented models, the GA-based model takes as input 
a computed feature vector for every 16-th note (i.e., 1/4 beats). Thus, the melody representation has a temporal resolution 8 times that of the chord progression (i.e., $N=8M$). This means that $\mathbf{x}_{\textcolor{black}{8}m}$ and $y_{m}$ point to the same temporal position. 

Our model uses a probabilistic approach, determining a fitness function based on the following elements. First,  the (logarithmic) conditional probability of the chord progression given the melody is represented as: 
    \begin{equation}
    F_1(X, Y) = \sum_{\textcolor{black}{n}=1}^{N} \log P(y_{\lceil \textcolor{black}{n/8} \rceil} | \mathbf{x}_{\textcolor{black}{n}}),
    \end{equation}
    where $\lceil~~\rceil$ is the ceiling function. 
    The chord transition probability is computed as: 
    \begin{equation}
    F_2(Y) = \sum_{m=3}^M \log P(y_m | y_{m-2}, y_{m-1}) \,.
    \end{equation}
    The conditional probability of each chord given its temporal position is defined as:
    \begin{equation}
    F_3(Y) = \sum_{m=1}^M \log P(y_m | {\rm Pos}_m) \,, 
    \end{equation}
    where ${\rm Pos}_m$ is the temporal position of the chord $y_m$. For simplicity, we defined ${\rm Pos}_m = {\rm mod}(m, 8)$, where ${\rm mod}$ is the modulo function. With this term, the model may learn that the tonic chord tends to appear at the first half of the first bar, while the dominant ($V$) chord tends to occur at the second half of the second bar.
    
    Finally, 
    we use the entropy to evaluate a chord sequence's complexity, which should not be too low as to avoid monotonous chord sequences.
    The entropy 
    is defined as $E(Y) = - \sum_{c_i \in {\cal C}} P(Y = c_i) \log P(Y = c_i)$.
    In the fitness function, we evaluate how likely this entropy $E(Y)$ is in a given data set. 
    \begin{equation}
    F_4(Y) = \log P(E = E(Y)) \,,    
    \label{eq:ga_entropy}
    \end{equation}
    where $E$ is the random variable of the entropy of chord progressions and is discritized by 0.25. Its probability distribution is obtained from the training data. 

The fitness function $F(Y)$ is calculated as: 
\begin{equation}
F(Y) = w_1 F_1(X, Y) + w_2 F_2(Y) + w_3 F_3(Y) + w_4 F_4(Y) \,.
\end{equation}
We simply set all the weights $w_1, w_2, w_3, w_4$ to 1.0 here.



\subsection{Deep BiLSTM-based Model}
\label{sec:bilstm}

This first deep learning model is adapted from the one proposed by \citep{lim17},
which uses BiLSTM layers.
This model extracts contextual information from the melody sequentially from both the positive 
and negative time directions. 
The original model makes chord prediction for every bar, using a  vocabulary of only the major and minor triad chords (i.e., $|\mathcal{C}|=24$).
We slightly extend this model such that the harmonic rhythm is a half  bar, and the output chord vocabulary includes diminished and augmented chords, and the N.C. symbol 
(i.e., $|\mathcal{C}|=49$). 

As shown in Figure \ref{fig:BiLSTM_Model}, this model has two BiLSTM layers, followed by a fully-connected layer. 
Dropout \citep{dropout} is applied with probability 0.2 at the output layer.
This dropout rate, as well as the number of hidden layers and hidden units, are empirically chosen by maximizing the chord prediction accuracy on a random held-out subset of the training set.
We train the model using minibatch gradient descent with categorical cross entropy as the the cost function. We use Adam 
as the optimizer and regularize by early stopping at the 10-th epoch to prevent over-fitting. 

\subsection{Deep Multitask Model: MTHarmonizer}
\label{sec:models:ours}

From our empirical observation on the samples generated by the aforementioned BiLSTM model, we find that the model has two main defects for longer phrases: 
\begin{enumerate}
\item \textbf{overuse of common chords}---common chords like C, F, and G major are repeated and overused, making the chord progression monotonous. 
\item \textbf{incorrect phrasing}---non-congruent phrasing between the melody and chords similarly results from the frequent occurrence of common chords.
The resulting frequent occurrence of progressions like F$\rightarrow$C or G$\rightarrow$C in generated sequences implies a musical cadence in an unfit location, potentially bringing an unnecessary sense of ending in the middle of a chord sequence. 
\end{enumerate}

\begin{figure}[]
    \centering
    \includegraphics[width=.6\columnwidth]{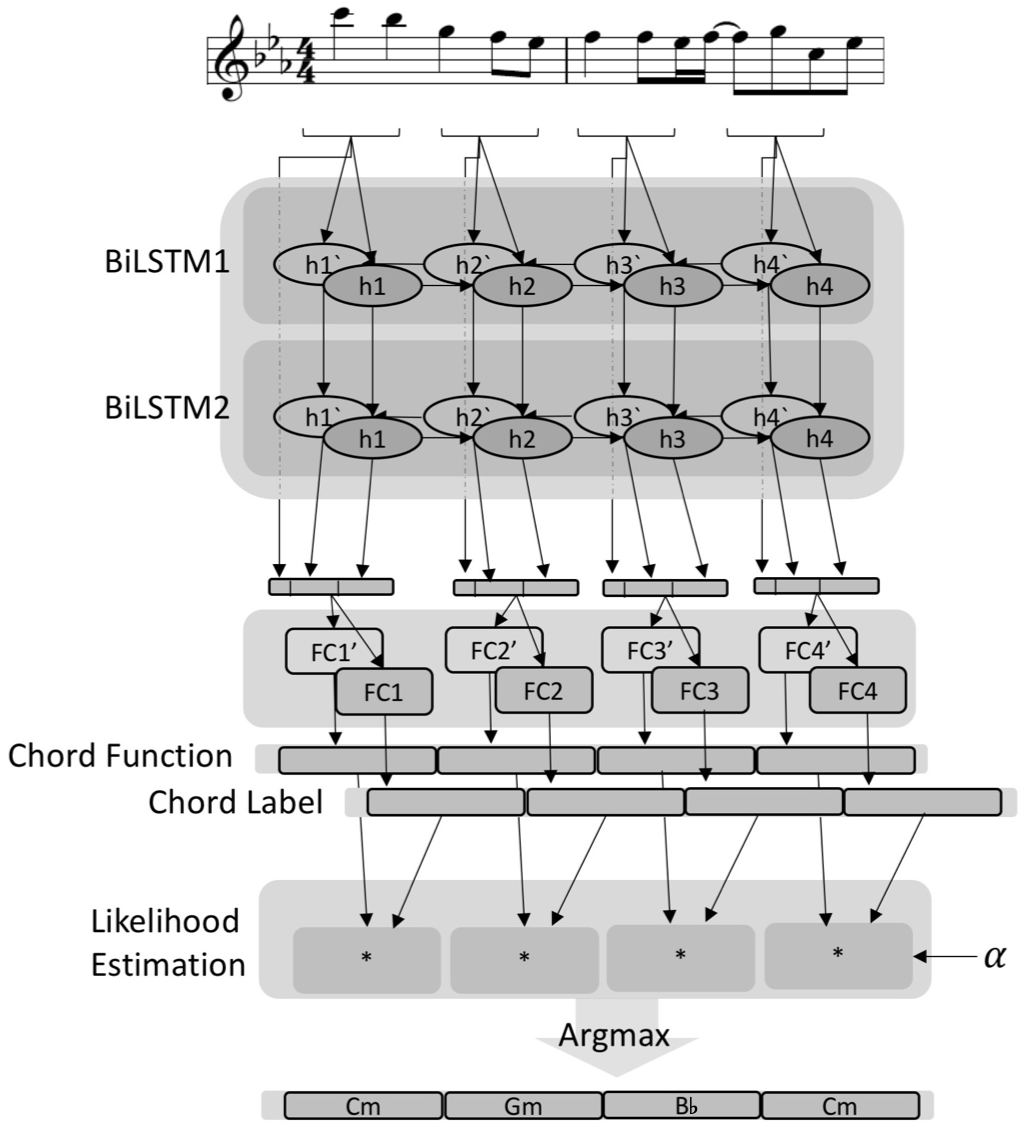}
    \caption{Diagram of the proposed MTHarmonizer, a deep multitask model extended from the model \citep{lim17} depicted in Figure \ref{fig:BiLSTM_Model}. See Section \ref{sec:models:ours} for details. }
    \label{fig:MTHarmonizer_Model}
\end{figure}

We propose an extension of the BiLSTM model to address these two defects. 
The core idea is to train the model to predict not only the chord labels but also the \emph{chord functions} \citep{chen18ismir},
as illustrated in Figure \ref{fig:MTHarmonizer_Model}. We call the resulting model a deep multitask model, or \emph{MTHarmonizer}, since it deals with two tasks at the same time. We note that the use of the chord functions for melody harmonization has been found useful by \citet{tsushima2018generative}, using an HMM-based model.


Functional harmony elaborates the relationship between chords and scales, and describes how harmonic motion guides musical perception and emotion \citep{chen18ismir}.
While a chord progression consisting of randomly selected chords generally feels aimless, chord progressions which follow the rules of functional harmony establish or contradict a tonality. 
Music theorists annotate each scale degree 
into functions, such as tonic and dominant,  
based on the association between chord and degree in a particular scale. These functions explain the role a given scale degree, and its associated chord relative to the scale, plays in musical phrasing and composition. Specifically, we consider the following functions:
\begin{itemize}
    \item The  \emph{tonic} function serves to stabilize and reinforce the tonal center.
    \item The \emph{dominant} function provides a strong sense of motion back to tonal center. 
    For example, a progression that moves from a dominant function scale degree chord to a tonal scale degree chord first creates tension, then resolves it.
    \item The \emph{others} that encompasses all the other chords that are neither tonic nor dominant, such as the subdominant chords.
\end{itemize}

As will be introduced in Section \ref{sec:dataset}, all the pieces in HTPD3 are in either C Major or c minor. Therefore, all chords share the same tonal center. We can directly map the chords into `tonic,' `dominant,' and `others' functional groups, by name, without worrying about their relative functions in other keys, for other tonal centers. 
Specifically, we consider C, Am, Cm, A as \emph{tonic} chords, and G and B diminished as \emph{dominant} chords. The other chords all fall into the \emph{others} category.

We identify two potential benefits of adding chord functions to the target output.
First, in contrast to the distribution of chord labels, the distribution of chord functions is relatively balanced, making it easier for the model to learn the chord functions.
Second, as the chord functions and chord labels are interdependent, adding the chord functions as a target informs the model which chord labels share the same function and may therefore be interchangeable.
We hypothesize that this multi-task learning will help our model learn proper functional progression, which in turn will produce better harmonic phrasing relative to the melody. 
Specifically, the loss function 
is defined as:
\begin{eqnarray}
L_{*} &=& L_\text{chord} + \gamma  L_\text{function} \nonumber \\
&=& H(\hat{Y}_\text{chord}, Y_\text{chord}) + \gamma  H(\hat{Y}_\text{function},Y_\text{function})  \nonumber \\
&=& H(f(X), Y_\text{chord}) + \gamma  H(g(X),Y_\text{function}) \,, 
\end{eqnarray}
where $H(\cdot)$ denotes the categorical cross entropy function, $f(\cdot)$ the chord label prediction branch, and $g(\cdot)$ the chord function prediction branch. 
When $\gamma=0$, the model reduces to the uni-task model proposed by \citet{lim17}, and we can simply write $Y_\text{chord}$ as $Y$.
In our work, we set 
$\gamma=1.5$ to ensure the loss value from $L_\text{chord}$ and $L_\text{function}$ are equally scaled. 
The two branches $f$ and $g$ share the two BiLSTM layers but not the fully-connected layer.
Empirically, we found that if $\gamma$ is too small, the model will tend to harmonize the melody with the chords with tonic and dominant functions; the resulting chord sequences would therefore lack diversity.

The outputs of $f$ and $g$ are likelihood values for each chord label and chord function given an input melody. As Figure \ref{fig:MTHarmonizer_Model} shows, in predicting the final chord sequence, we rely on a weighted combination of the outputs of $f$ and $g$ in the following way: 
\begin{equation}
\hat{Y} = \argmax_{\hat{y}_1,\hat{y}_2,\dots,\hat{y}_M} \prod_{m=1}^M (  P(\hat{y}_m = f(\mathbf{x}_m)) * \textcolor{black}{\alpha_m} P(\hat{y}_m = h(g(\mathbf{x}_m))) ) \,,
\end{equation}
where 
$h(\cdot)$ is simply a look-up table that maps the three chord functions to the $|\mathcal{C}|$ chord labels, and \textcolor{black}{$\alpha_m$} is a pre-defined hyperparameter that allows us to boost the importance of correctly predicting the chord function over that of correctly predicting the chord label, for each chord. In our implementation, we set
$\alpha_m=1.0$ for the tonic and dominant chords, and 
$\alpha_m=1.8$ for the other chords, to  encourage the model to select chord labels that have lower likelihood, i.e., to use the ``others'' chords. 
This would more likely affect the middle part of a chord sequence, because this is where the likelihood to observe a chord from the three functions to be likely similar, so applying different $\alpha_m$ makes a difference. 
In contrast, in the beginning or the end of a phrase, the likelihood of observing the ``others'' chords would tend to be low anyway, even after we boost it with $\alpha_m$.
As we will mainly add diversity to the middle part of a chord sequence, we would not compromise the overall chord progression and phrasing.

\section{Proposed Dataset}\label{sec:dataset}



For the purpose of this study, we firstly collect a new dataset called the Hooktheory Lead Sheet Dataset (HLSD), which consists of lead sheet samples scraped from the online music theory forum called TheoryTab, hosted by Hooktheory (\url{https://www.hooktheory.com/theorytab}), a company that produces pedagogical music software and books. The majority of lead sheet samples found on TheoryTab are user-contributed.\footnote{We note that we do not own the copyrights of the lead sheets so we cannot further redistribute them. We collected the lead sheets only for academic research.} Each piece contains high-quality, human-transcribed melodies alongside their corresponding chord progressions, which are specified by both literal chord symbols (e.g., Gmaj7), and chord functions (e.g., VI7) relative to the provided key.Chord symbols specify inversion if applicable, and the full set of chord extensions (e.g., \#9, b11). The metric timing/placement of the chords is also provided.
Due to copyright concerns, TheoryTab prohibits uploading full length songs. Instead, users upload snippets of a song (here referred to as lead sheet samples), which they voluntarily annotate with structural labels (e.g. ``Intro," ``Verse," and ``Chorus") and genre labels. A music piece can be associated with multiple genres.

As the samples in this dataset are segments of pop songs, e.g., a verse or a chorus, temporary changes of the tonal center should be rare.

We note that HLSD contains music of various genres, and only a few of them are classical music.\footnote{ Here is a random sample of 10 songs from the dataset: \emph{Love Grows Where My Rosemary Goes} by Edison Lighthouse (1972),
\emph{La Bamba} by Ritchie Valen (1987),
\emph{Palm Tree Paradise-Wario Land 4} by Ryoji Yoshitomi (2001),
\emph{Forever and Always} by Taylor Swift (2008),
\emph{Trails of the Past} by Sbtrkt (2011),
\emph{I’ve Run Away to Join the Fairies} by Magnetic Fields (2012),
\emph{Semi Automatic} by Twenty One Pilots (2013),
\emph{Adventure of A Lifetime} by Coldplay (2015),
\emph{Same Drugs} by Chance the Rapper (2016),
\emph{Feel Good-Brooks Remix} by Gryffin And Illenium (2017).}
As discussed in \citep{declercq_temperley_2011}, the rules of classical harmony
are much less often followed in pop music. In addition, melodies in pop/rock music are more independent of the harmony than the case in classical music \citep{temperley_2007,nobile15}. 
It therefore remains to be studied whether the consideration of functional harmony improves our task here.

HLSD contains 11,329 lead sheets samples, all in 4/4 time signature. 
It contains up to 704 different chord classes, which is deemed too many for the current study. 
We therefore take the following steps to process and simplify HLSD, resulting in the final HTPD3 dataset employed in the performance study.

\begin{itemize}
\item We remove lead sheet samples that do not contain a sufficient number of notes. Specifically, we remove samples whose 
melodies comprise of more than 40\% rests (relative to their lengths). One can think of this as correcting class imbalance, another common issue for machine learning models---if the model sees too much of a single event, it may overfit and only produce or classify that event. 
\item We then filter out lead sheets that are less than 4 bars and longer than 32 bars, so that $4\leq T \leq 32$. 
This is done because 4 bars is commonly seen as the minimum length for a complete musical phrase in 4/4 time
signature. At the other end, 32 bars is a common length for a full lead sheet, one that is relatively long. 
Hence, as the majority of our dataset consists of mere song sections, we are inclined for not including samples longer than 32 bars.
\item The HLSD provides the key signatures of every samples. We transpose every samples to either C major or c minor based on the provided key signatures.
\item In general, a chord label can be specified by the pitch class of its root note (among 12 possible pitch classes, i.e., \texttt{C}, \texttt{C\#}, $\dots$, \texttt{B}, in a chromatic scale), and its chord quality, such as `triad', `sixths', `sevenths', and `suspended.' HLSD contains 704 possible chord labels, including inversions. However, the distribution of these labels is highly skewed. 
In order to even out the distribution and simplify our task, we reduce the chord vocabulary by converting each label to its root position triad form, i.e., the major, minor, diminished, and augmented chords without 7ths or additional extensions. Suspended chords are mapped to the major and minor chords.
As a result, only 48 chord labels (i.e., 12 root notes by 4 qualities) and N.C. are considered (i.e., $|\mathcal{C}|=49$). 
\item We standardize the dataset so that a chord change can occur only every bar or every half bar. 
\end{itemize}

We do admit that this simplification can decrease the chord color and reduce the intensity of tension/release patterns, and 
can sometimes convert a vibrant, subtle progression into a monotonous one (e.g., because both CMaj7 and C7 are mapped to C chord). 
We plan to make full use of the original chord vocabulary in future works.


Having pre-defined train and test splits helps to facilitate the use of HTPD3 for evaluating new models of melody harmonization via the standardization of training procedure. As HTPD3 includes paired melody and chord sequences, it can also be used to evaluate models for chord-conditioned melody
generation as well.
With these use cases in mind, we split the dataset so that the \textbf{training set} contains 80\% of the pieces, and the \textbf{test set} contains 10\% of the pieces.
There are in total 923 lead sheet samples in the test set. 
The remaining 10\% is reserved for future use.
When splitting, we imposed the additional requirement that lead sheet samples from the same song are in the same subset.

\section{Proposed Objective Metrics}
\label{sec:metrics}


To our knowledge, there are at present no standardized, objective evaluation metrics for the melody harmonization task. The only objective metric adopted by \citep{lim17}, in evaluating the models they built is a categorical cross entropy-based chord prediction error, representing the discrepancy between the ground truth chords $Y_*$ 
and predicted chords $\hat{Y}_*=f(X_*)$.
The chord prediction error is calculated for each half bar individually and then got averaged, not considering the chord sequence as a whole. In addition, it does not directly measure how the generated chord sequence 
fits with the given melody. 
What's more, when calculating the chord prediction error, the underlying assumption is that the ``ground truth'' chord sequence $Y_*$ is the \emph{only} feasible one to harmonize the given melody $X_*$. This is not true in general.

For the comparative study, we introduce here a set of six objective metrics defined below.
These metrics are split into two categories, namely three chord progression metrics and three chord/melody harmonicity metrics.
Please note that we do not evaluate the melody itself, as the melody is provided by the ground truth data.

\emph{Chord progression metrics}  evaluate each chord sequence as a whole, independent from the melody, and relate to the distribution of chord labels in a sequence. 
\begin{itemize}
  \item \textbf{Chord histogram entropy} (CHE):
  Given a chord sequence, we create a histogram of chord occurrences with $|\mathcal{C}|$ bins. Then, we normalize the counts to sum to 1, and calculate its entropy:
  \begin{equation}
    H = -\sum_{i=1}^{|\mathcal{C}|}p_{i}\log{p_i }\,,
  \end{equation}
  where $p_i$ is the relative probability of the $i$-th bin. The entropy is greatest when the histogram follows a uniform distribution, and lowest when the chord sequence uses only one chord throughout.
  \item \textbf{Chord coverage} (CC): 
  The number of chord labels with non-zero counts in the chord histogram in a chord sequence.
  \item \textbf{Chord tonal distance} (CTD):
  The \emph{tonal distance} proposed by \citep{tonalDist} is a canonical way to measure the closeness of two chords. It is calculated 
  \textcolor{black}{by firstly calculating the PCP features of two chords, projecting the PCP features to a derived 6-D tonal space, and finally calculating the Euclidean distance between the two 6-D feature vectors}.
  CTD is the average value of the tonal distance computed between every pair of adjacent chords in a given chord sequence. The CTD is highest when there are abrupt changes in the chord progression (e.g., from C chord to B chord).
  
\end{itemize}

\emph{Chord/melody harmonicity metrics}, on the other hand, aims to evaluate the degree to which a generated chord sequence successfully harmonizes a given melody sequence. 
\begin{itemize}
  \item \textbf{Chord tone to non-chord tone ratio} (CTnCTR):
    In reference to the chord sequence, we count the number of \emph{chord tones}, and \emph{non-chord tones} in the melody sequence. Chord tones are defined as melody notes whose pitch class are part of the current chord (i.e., one of the three pitch classes that make up a triad) for the corresponding half bar.
    All the other melody notes are viewed as non-chord tones. 
    One way to measure the harmonicity is to simply computing the ratio of the number of the chord tones ($n_c$) to the number of the non-chord tones ($n_n$). 
    However, we find it useful to further take into account the number of a subset of non-chord tones ($n_p$) 
    that are two semitones within the notes which are right after them, where subscript {\it p} denotes a ``proper'' non-chord tone. We define CTnCTR as
    \begin{equation}
        \frac{n_c+n_p}{n_c+n_n}.
    \end{equation}
    CTnCTR equals one when there are no non-chord tones at all, or when $n_p=n_n$.

    \item \textbf{Pitch consonance score} (PCS): For each melody note, we calculate a \emph{consonance score} with each of the three notes of its corresponding chord label.
   The consonance scores are computed based on the musical interval between the pitch of the melody notes and the  chord notes, assuming that the pitch of the melody notes is always higher. This is always the case in our implementation, because we always place the chord notes lower than the melody notes.
   The consonance score is set to 1 for consonance intervals including unison, major/minor 3rd, perfect 5th, major/minor 6th, set to 0 for a perfect 4th, and set to --1 for other intervals, which are considered dissonant. 
   PCS for a pair of melody and chord sequences is computed by averaging these consonance scores across a 16th-note windows, excluding rest periods. 
   
\item \textbf{Melody-chord tonal distance} (MCTD): 
Extending the idea of tonal distance, we represent a melody note by a PCP feature vector (which would be a one-hot vector) and compare it against the PCP of a chord label in the 6-D tonal space  \citep{tonalDist} to calculate the closeness between a melody note and a chord label.
MCTD is the average of the tonal distance between every melody note and corresponding the chord label calculated across a melody sequence, with each distance weighted by the duration of the corresponding melody note.
\end{itemize}

\section{Comparative Study}\label{sec:eval}

\begin{figure}[]
    \centering
    \includegraphics[width=\columnwidth]{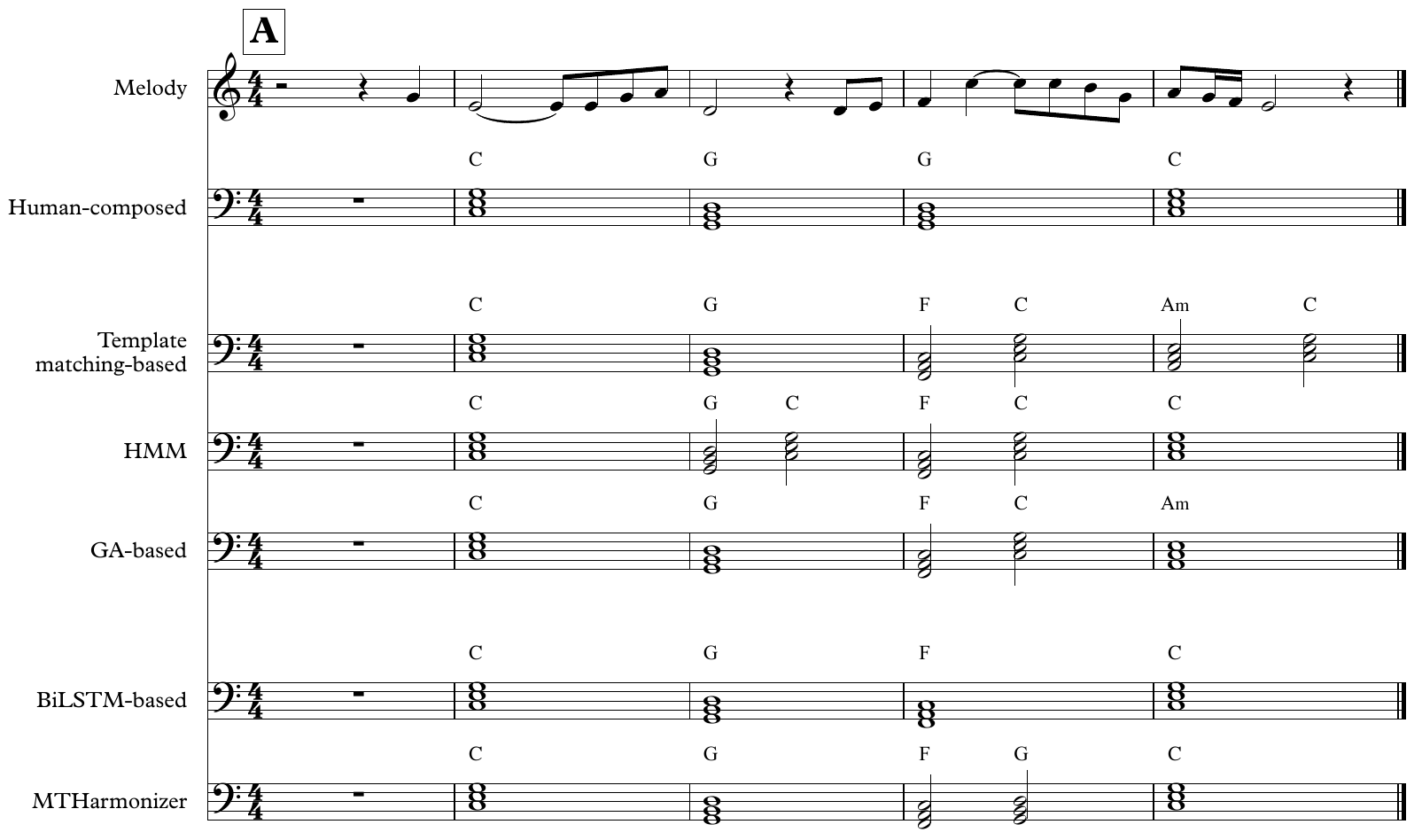}
    \caption{A harmonization example (in major key) from The Beatles: \textit{Hey Jude}. We can see that, while the non-deep learning models change the harmonization in different phrases, the  MTharmonizer generates a V-I progression nicely to close the phrase.}
    \label{fig:HeyJude_sheet}
\end{figure}

\begin{figure}[]
    \centering
    \includegraphics[width=\columnwidth]{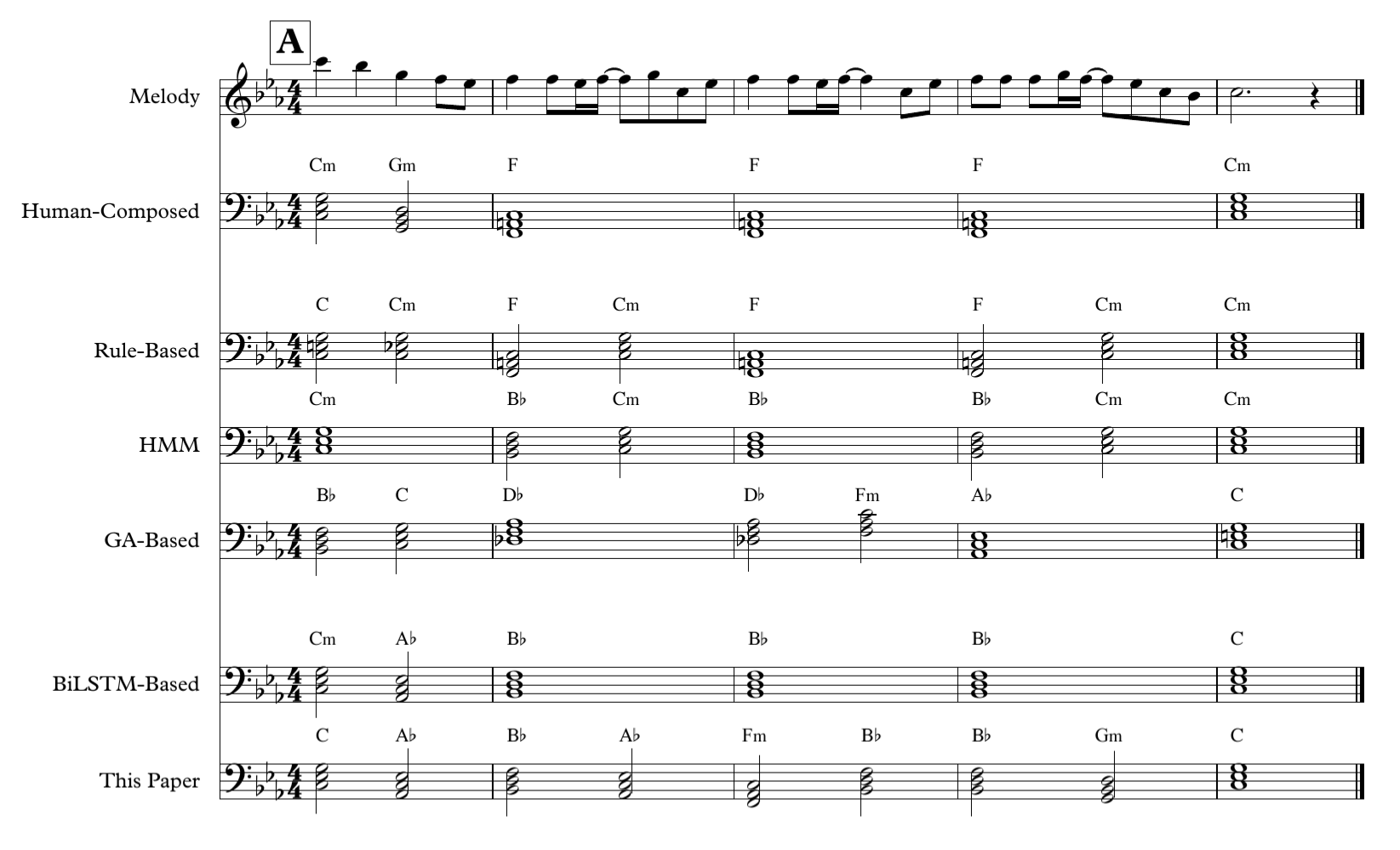}
    \caption{A harmonization example (in minor key) from ABBA: \textit{Gimme Gimme Gimme A Man After Midnight}. Similar to the example shown in Figure \ref{fig:HeyJude_sheet}, the result of the MTHarmonizer appears to be more diverse and functionally correct. We also see that the result of GA is quite ``interesting''---e.g., with non-diatonic chord D flat Major and close the music phrase with Picardy third (i.e., a major chord of the tonic at the end of a chord sequence that is in a minor key). We also see that the non-deep learning methods seem to be weaker in handling the tonality of music.}
    \label{fig:GimmeMidnight_sheet}
\end{figure}

We train all the five models described in Section \ref{sec:models} using the training split of HTPD3 and then apply them to the test split of HTPD3 to get the predicted chord sequences for each melody sequence.
Examples of the harmonization result of the evaluated models can be found in 
Figures \ref{fig:HeyJude_sheet} and \ref{fig:GimmeMidnight_sheet}.

We note that, since one cannot judge the full potential of each algorithm only from our simplified setting of melody harmonization, we do not intend to find what method is the best in general. We rather attempt a challenge to compare different harmonization method which have not been directly compared because of the different context that each approach assumes.

In what follows, we use the harmonization result for a random subset of the test set comprising 100 pieces in a user study for subjective evaluation. The result of this subjective evaluation is presented in Section \ref{sec:subeval}. 
Then, in Section \ref{sec:objeval}, we report the results of an objective evaluation wherein we compute the mean values of the chord/melody harmonicity and chord progression metrics presented in Section \ref{sec:metrics} for the harmonization results for each test set piece.

\subsection{Subjective Evaluation}\label{sec:subeval}

We conducted an online survey where we invited human subjects to listen to and assess the harmonization results of different models. The subjects evaluated the harmonizations in terms of the following criteria:
\begin{itemize}
    \item \textbf{Harmonicity}: The extent to which a chord progression successfully or pleasantly harmonizes a given melody. This is designed to correspond to what the melody/chord harmonicity metrics described in Section \ref{sec:metrics} aim to measure.
    \item \textbf{Interestingness}: The extent to which a chord progression sounds exciting, unexpected and/or generates ``positive” stimulation. This criterion corresponds to the chord-related metrics described in Section \ref{sec:metrics}. Please note that we use a less technical term ``interestingness'' here since we intend to solicit feedback from people either with or without musical backgrounds. \item The \textbf{Overall} quality of the given harmonization.
\end{itemize}

Given a melody sequence, we have in total six candidate chord sequences to accompany it: those generated by the five models presented in Section \ref{sec:models}, and the \emph{human-composed}, ground-truth progression retrieved directly from the test set. We intend to compare the results of the automatically generated progression with the original human-composed progression. Yet, given the time and cognitive load required, it was not possible to ask each subject to evaluate the results of every model for every piece of music in the test set (there are $6\times923=5,538$ sequences in total). 
We describe below how our user study is designed to make the evaluation feasible. 


\subsubsection{Design of the User Study}\label{sec:subeval_design}

First, we randomly select 100 melodies from the test set of HTPD3. For each human subject, we randomly select three melody sequences from this pool, and present to the subject the harmonization results of two randomly selected models for each melody sequence. For each of the three melodies, the subject listens to the melody without accompaniment first, and then the sequence with two different harmonizations. Thus, the subject has to listen to nine music pieces in total: three melody sequences and the six harmonized ones. As we have six methods for melody harmonization (including the original human-composed harmonization), we select methods for each set of music 
such that each method is presented once and only once to each subject. The subjects are not aware of which harmonization is generated by which method, but are informed that at least one of the harmonized sequence is human-composed.

In each set, the subject has to listen to the two harmonized sequences and decide which version is better according to the three criteria mentioned earlier. This \emph{ranking} task is mandatory. In addition, the subject can choose to further grade the harmonized sequences in a five-point Likert scale with respect to the criteria mentioned earlier. Here, we break ``harmonicity'' into the following two criteria in order to get more feedback from subjects:
\begin{itemize}
    \item \textbf{Coherence}: the coherence between the melody and the chord progression in terms of harmonicity and phrasing. 
    \item \textbf{Chord Progression}: how coherent, pleasant, or reasonable the chord progression is on its own, independent of the melody.
\end{itemize}
This optional \emph{rating} task thus has four criteria in total.

The user study opens to an ``instructions'' page, that informs the subjects that we consider only  root-positioned triad chords in the survey. Moreover, they are informed that there is no ``ground truth” in melody harmonization---the task is by nature subjective.
After collecting a small amount of relevant personal information from the subjects, we present them with a random audio sample and encourage them to put on their headsets and 
adjust the volume to a comfortable level. After that, they are prompted to begin evaluating the three sets (i.e., one set for each melody sequence), one-by-one on consecutive pages.

We spread the online survey over the Internet openly, without restriction, to solicit voluntary, non-paid participation.
The webpage of the survey can be found at \url{https://musicai.citi.sinica.edu.tw/survey_mel_harm/}.


\begin{figure}[]
  \centering
  \begin{tabular}{cc}
    \includegraphics[width=.48\columnwidth]{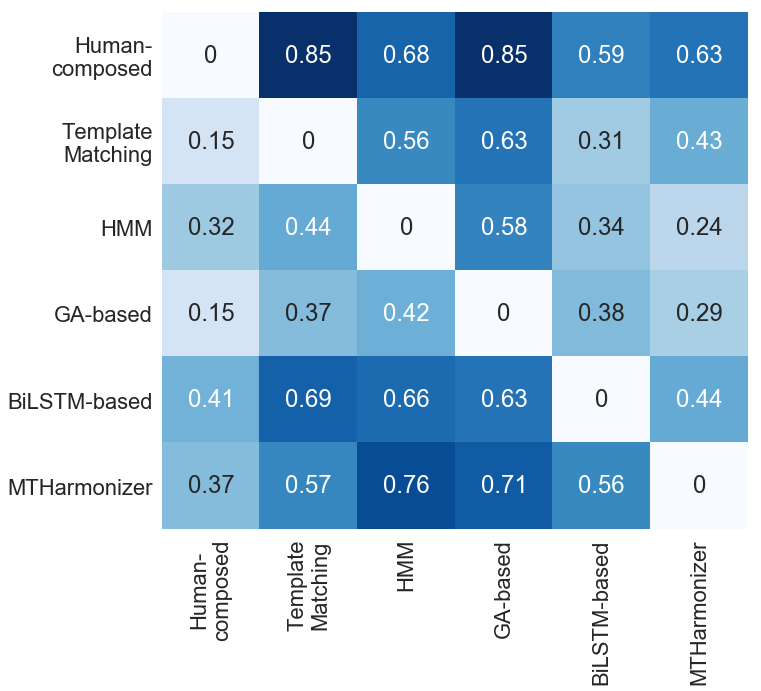}
    &
    \includegraphics[width=.48\columnwidth]{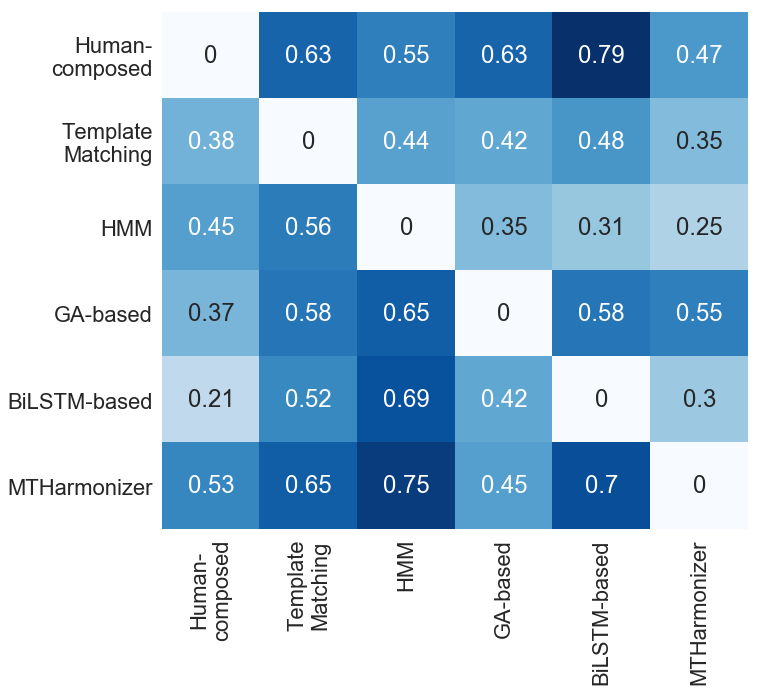} \\
    (a) Harmonicities &(b) Interestingness 
  \end{tabular} 
  \includegraphics[width=.48\columnwidth]{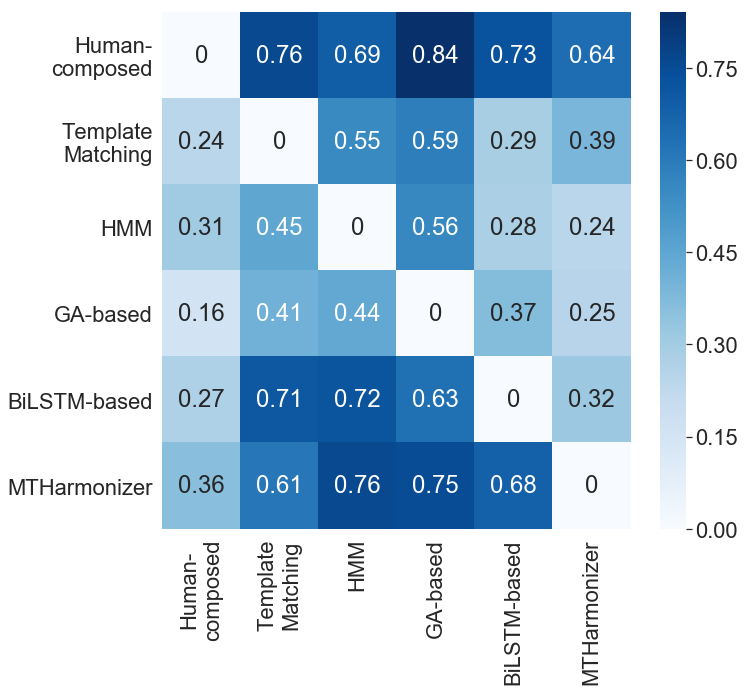}\\
    (c) Overall
  \caption{``Win probabilities'' of different model pairs. Each entry represents the probability that the model in that column scores higher than the model in that row.}
  \label{fig:WinProb}
\end{figure}


\subsubsection{User Study Results}
\label{sec:subeval_result}
In total, 202 participants from 16 countries took part in the survey. 
We had more male participants than female (ratio of 1.82:1), and the average age of participants was 30.8 years old. 
122 participants indicated that they have music background, and 69 of them are familiar with or expertise in the harmonic theory. 
The participants took on average 14.2 minutes to complete the survey. 

We performed the following two data cleaning steps: 
First, we discarded both the ranking and rating results from participants who spent less than 3 minutes to complete the survey, which is considered too short. 
Second, we disregarded rating results when the relative ordering of the methods contradicted that from the ranking results. As a result, 9.1\% and 21\% of the ranking and rating records were removed, respectively. 

We first discuss the results of the pairwise ranking task, which is shown in Figure \ref{fig:WinProb}. The following observations are made:
\begin{itemize}
    \item The human-composed progressions have the highest ``win probabilities'' on average in all the three ranking criteria. It performs particularly well in \emph{Harmonicity}.
    \item In general, the deep learning methods have higher probabilities to win over the non-deep learning methods in \emph{Harmonicity} and \emph{Overall}. 
    \item For \emph{Interestingness}, GA performs the best among the five automatic methods, which we suspect stems from its entropy term (Eq. (\ref{eq:ga_entropy})).
    \item Among the two deep learning methods, the MTHarmonizer consistently outperforms the  BiLSTM in all  ranking criteria, especially for \emph{Interestingness}. We (subjectively) observe that MTHarmonizer indeed generates more diverse chord progressions compared to the vanilla BiLSTM, perhaps due to the consideration of functions.
\end{itemize}


\begin{figure*}[]
    \centering
    \includegraphics[width=\columnwidth]{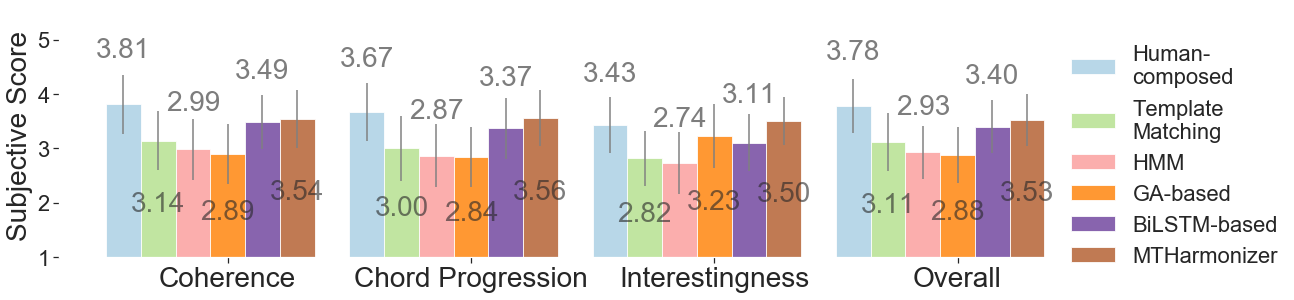}
    \caption{The mean rating scores in subjective evaluation, along with the standard deviation  (the error bars).}
    \label{fig:EvalScore}
\end{figure*}

The results of the rating task shown in Figure \ref{fig:EvalScore}, on the other hand, lead to the following observations:
\begin{itemize}
    \item Congruent with the results of the ranking task, the MTHarmonzer model achieves the second best performance here, only losing out to the original human-composed chord progressions. The MTHarmonzier consistently outperforms the other four automatic methods in all the four metrics. With a paired t-test,
    we find that there is significant performance difference between the MTHarmonzer progressions and the original  human-composed progressions in terms of \emph{Coherence} and \emph{Chord Progression} (p-value$<$0.005), but no significant difference in terms of \emph{Interestingness} and  \emph{Overall}.
    \item Among the four metrics, the original human-composed progressions score higher in \emph{Coherence} (3.81) and \emph{Overall} (3.78), and the lowest in \emph{Interestingness} (3.43). This suggests that the way we simplify the data (e.g., using only root-positioned triad chords) may have limited the perceptual qualities of the music, in particular its diversity. 
    
    \item Generally speaking, the results in \emph{Chord Progression} (i.e., the coherence of the chord progression on its own) seems to correlate better with the results in \emph{Coherence} (i.e., the coherence between the melody and chord sequences) than the \emph{Interestingness} of the chord progression. This suggests that a chord progression rated as being interesting may not sound coherent.
    
    \item Although the GA performs worse than the MTHarmonizer on all the four metrics, it actually performs fairly well in \emph{Interestingness} (3.23), as we have observed from the ranking result. A paired t-test showed no significant performance difference between the GA generated progressions and original human-composed progressions in \emph{Interestingness}. A hybrid model that combines GA and deep learning may be a promising direction for future research.
\end{itemize}

\begin{table}
  \centering
  \begin{tabular}{l ccc}
    \toprule
    \textbf{Melody/chord harmonicity metrics}    &CTnCTR        & PCS          &MCTD\\ 
    \midrule
    Human-composed                               &0.74          &1.42          &1.03\\[1ex]
    Template matching                            &0.91          &1.97          &0.83\\ 
    HMM (adapted from \citep{lim17})             &0.89          &1.93          &0.85\\
    GA-based (adapted from \citep{kitahara2018}) &\textbf{0.74} &0.43          &1.31\\
    BiLSTM (adapted from \citep{lim17})          &0.87          &1.84          &0.91\\
    MTHarmonizer (proposed here)                 &0.82          &\textbf{1.77} &\textbf{0.94}\\
    \midrule
    \textbf{Chord progression metrics}           &CHE           &CC            &CTD\\
    \midrule
    Human-composed                               &1.28          &2.62          &0.88\\[1ex]
    Template matching                            &1.01          &1.70          &0.65\\ 
    HMM (adapted from \citep{lim17})             &0.88          &1.89          &0.56\\
    GA-based (adapted from \citep{kitahara2018}) &1.58          &\textbf{2.47} &\textbf{0.96}\\
    BiLSTM (adapted from \citep{lim17})          &1.07          &2.07          &0.71\\
    MTHarmonizer (proposed here)                 &\textbf{1.29} &2.31          &1.02\\
    \bottomrule
  \end{tabular}
  \caption{Objective evaluation scores for different melody harmonization models. The closer the values to that of the human-composed samples, the better the model is in modeling the training data. The bold values indicate the closet value to that of the human-composed samples per metric. And, 1) higher values in CTnCTR and PCS and lower values in MCTD may suggest that the melody/chord harmonicity is high; 2) higher values in CHE and CC and lower values in CTD may suggest that the diversity (which can be related to the interestingness) of the chord progression is high. 
  See Section \ref{sec:metrics} for the definitions of the metrics.}
  \label{tab:evalscores}
\end{table}    

From the rating and ranking tasks, we see that, in terms of harmonicity, automatic methods still fall behind the human composition. However, the results of the two deep learning based methods are closer to that of the human-composed ones.  



\subsection{Objective Evaluation}\label{sec:objeval}




The results are displayed in Table \ref{tab:evalscores}. 
We discuss the result of the melody/chord harmonicity metrics first. 
We can see that the results for the two deep learning methods are in general closer to the results for the original human-composed progressions than those of the three non-deep learning methods for all three harmonicity metrics, most significantly on the latter two. 
The template matching-based and HMM-based methods scores high in PCS and low in MCTD, indicating that the harmonization these two methods generate may be too conservative. 
In contrast, the GA scores low in PCS and high in MCTD, indicating overly low harmonicity.
These results are consistent with the subjective evaluation, suggesting that these metrics can perhaps reflect human perception of the harmonicity between melody and chords. 

From the result of the chord progression metrics, we also see from CHE and CC that 
the progressions generated by the template matching-based and HMM-based methods seem to lack diversity. 
In contrast, the output of GA  features high diversity. 

As the GA based method was rated lower than the template matching and HMM methods in terms of the \emph{Overall} criterion in our subjective evaluation, it seems that the subjects care more about the harmonicity than the diversity of chord progressions.


Comparing the two deep learning methods, we see that the MTHarmonizer uses more non-chord tones (smaller CTnCTR) and uses a greater number of unique chords (larger CC) than the BiLSTM model. The CHE of the MTHarmonizer is very close to that of the original human-composed progressions.

In general, the results of the objective evaluation appear consistent with those of the subjective evaluation. It is difficult to quantify which metrics are better for what purposes, and how useful and accurate these metrics are overall. Therefore, our suggestion is to use them mainly to gain practical insights into the results of automatic melody harmonization models, rather than to judge their quality. 
As pointed out by \citep{musegan}, objective metrics can be used to track the performance of models during development, before committing to running the user study. Yet, human evaluations are still needed to evaluate the quality of the generated music.

Finally, although we have argued earlier that there is no ground truth in melody harmonization and that it is not adequate to rely solely on chord prediction error to evaluate the models, we find that the MTHarmonizer also achieves the lowest chord prediction error (i.e., a 48-class classification problem) and chord function prediction error (a 3-class classification problem). On the test set, the chord prediction accuracy for the five models (template matching, HMM, GA, BiLSTM, and MTHarmonizer) is 29\%, 31\%, 20\%, 35\%, and 38\%, respectively, while the accuracy for a random guess baseline is only 2\%. And, the chord function prediction accuracy is 62\%, 61\%, 55\%, 65\%, 69\%, respectively, while the accuracy for a random guess baseline is 51\%.



\section{Discussions}
\label{sec:discussion}

We admit that the comparative study presented above has some limitations. First, because of the various preprocessing steps taken for data cleaning and for making the melody harmonization task manageable (cf. Section \ref{sec:dataset}), the ``human-composed'' harmonizations are actually simplified versions of those found on TheoryTab. We considered triad chords only, and we did not consider performance-level attributes such as velocity and rhythmic pattern of chords.
This limits the perceptual quality of the human-composed chord progression, and therefore also limits the results that can be achieved by automatic methods. 
The reduction from extended chords to triads reduces the ``color'' of the chords and creates many innacurate chord repetitions in the dataset (e.g., both the alternated CMaj7 and C7 will be reduced to C triad chord). 
We believe it is important to properly inform the human subjects of such limitations as we did in the instruction phase of our user study.
We plan to compile other datasets from HLSD to extend the comparative study in the future.


Second, in our user study we asked human subjects to rank and rate the results of two randomly chosen methods in each of the three presented sets. After analyzing the results, we found that the subject's ratings are in fact \emph{relative}. For example, the MTHarmonizer's average score in \emph{Overall} is 3.04 when presented alongside the human-composed progressions, and 3.57 when confronted with the genetic algorithm-based model. 
We made sure in our user study that all the methods are equally likely to be presented together with every other method, so the average rating scores presented in Figure \ref{fig:EvalScore}  do not favor a particular method. Still, caution is needed when interpreting the rating scores.
Humans may not have a clear idea of how to consistently assign a score to a harmonization. 
While it is certainly easier to objectively compare multiple methods with the provided rating scores, we still recommended asking human subjects to make pairwise rankings in order to make the result more reliable.

Third, as the aim of this paper is to sample representative models from the literature, the current setting cannot decouple the effects of the model architectures (e.g., HMMs, GAs and BiLSTMs) and the music knowledge induced into the model (e.g., the concept of functional harmonic in MTHarmonizer).
Moreover, for similar reasons, the models we implemented in this paper do not include extensions that might lead to better performance. For example, we could further improve the HMM model by using trigrams or extending the hidden layers as discussed in the literature~\citep{paiement2006pmh,temperley2009unified,tsushima17ismir}. 
It may also be possible to incorporate some of the proposed objective metrics (such as the chord tone to non-chord tone ratio) as  additional loss terms. 
The aim of this paper is to observe how different categories of models characterize the harmonization results rather than to explore the full potentials of each presented model.



Reviewing the properties of harmonization algorithms which imitate styles in a dataset as in our research still holds its importance, although recent music generation research is shifting towards measuring how systems can generate content that extrapolates meaningfully from what the model have learned \citep{zacharakis18me}. Extrapolation could be based on the model which also achieves interpolation or maintaining particular styles among data points. We believe we can further discuss extrapolation based on the understanding of how methods imitate data.

\section{Conclusion}\label{sec:conclusion}

In this paper, we have presented a comparative study implementing and evaluating a number of canonical methods and one new method for melody harmonization, including deep learning and non-deep learning based approaches. The evaluation has been done using a lead sheet dataset we newly collected for training and evaluating melody harmonization. 
In addition to conducting a subjective evaluation, we employed in total six objective metrics with which to evaluate a chord progression given a melody.
Our evaluation shows that deep learning models indeed perform better than non-deep learning ones in a variety of aspects, including  harmonicity and interestingness. Moreover, a deep learning model that takes the function of chords into account reaches the best result among the evaluated models. 


Future work can be directed toward at least the following three directions.
First, it is important to extend the chord vocabulary to include more complicated chords. 
Second, as self-attention based neural sequence models such as the Transformers \citep{vaswani2017attention} have been shown powerful alternatives to RNNs for various automatic music generation tasks \citep{huang2020pop, guitarTransformer20ismir, donahue2019lakhnes,ren20mm}, it might be interesting to investigate Transformer-based models for melody harmonization. Finally, other than the melody harmonization task (i.e., generating chords given a melody) addressed in the paper, it would also be interesting to study chord-conditioned melody generation (i.e., generating a melody given chords) \citep{midinet,jazzgan,genchel19csmc}, or simultaneously generating both melody and harmony from scratch \citep{liu18icmla,jiang20icassp,jazzTransformer20ismir}.

\bibliographystyle{apacite}
\bibliography{ref}

\end{document}